\documentclass[aps,prl,twocolumn,epsfig,amsmath,amssymb]{revtex4}
\usepackage[english]{babel} \usepackage{latexsym}
\usepackage{graphicx} \usepackage{subfigure} \usepackage{epsfig}
\usepackage{psfrag}

\begin{document}

\title{Scattering of two-dimensional solitons in dipolar Bose-Einstein 
condensates}
\author{R. Nath$^1$, P. Pedri$^2$ and L. Santos$^1$} 
\affiliation{
\mbox{$^1$Institut f\"ur Theoretische Physik , Leibniz Universit\"at
Hannover, Appelstr. 2, D-30167, Hannover, Germany}\\
\mbox{$^2$Laboratoire de Physique Th\'eorique et Mod\`eles Statistiques,
Universit\'e Paris Sud, 91405 Orsay Cedex, France}\\
}

\begin{abstract}  
%
%

We analyze the scattering of bright solitons in dipolar Bose-Einstein condensates placed 
in unconnected layers. Whereas for short-range interactions unconnected layers are independent, 
a remarkable consequence of the dipole interaction is the appearance  
of novel nonlocal interlayer effects. In particular, 
we show that the interlayer interaction leads to an effective molecular potential between disconnected solitons, inducing 
a complex scattering physics between them, which includes inelastic fusion into soliton-molecules, 
and strong symmetric and asymmetric inelastic resonances. In addition, 
a fundamentally new 2D scattering scenario in matter-wave solitons is possible, in which inelastic spiraling occurs, 
resembling phenomena in photorrefractive materials. Finally, we consider the scattering of unconnected 1D solitons 
and discuss the feasibility in current on going experiments.

\end{abstract}  
 \maketitle

Up to very recently, typical experiments on ultra cold gases involved 
particles interacting dominantly via a short-range isotropic  
potential, which, due to the very low energies involved, is 
fully determined by the corresponding $s$-wave scattering length. However, recent 
experiments on cold molecules \cite{Molecules}, Rydberg atoms \cite{Rydberg}, and 
atoms with large magnetic moment \cite{Chromium}, open a fascinating new 
research area, namely that of dipolar gases, for which the dipole-dipole 
interaction (DDI) plays a significant or even dominant role. The DDI 
is long-range and anisotropic (partially attractive), 
and leads to fundamentally new physics in condensates 
\cite{Stability,Excitations,Roton}, degenerated Fermi gases \cite{Fermions}, and strongly-correlated atomic systems 
\cite{DipLat-FQHE}. It leads to the Einstein-de Haas effect in spinor condensates \cite{EdH}, 
and may be employed for quantum computation \cite{QInf}, and ultra cold chemistry \cite{Chemistry}.
Recently, time-of-flight experiments in Chromium have allowed for the first observation ever  
of dipolar effects in quantum gases  \cite{Expansion}.

Interestingly, the physics of short-range interacting  
Bose-Einstein condensates (BECs) at low temperatures is given by a nonlinear Schr\"odinger equation (NLSE) with 
cubic local nonlinearity, similar to the one appearing in other systems, in particular in 
Kerr media in nonlinear optics. In 1D, nonlinearity allows for solitonic solutions \cite{Zakharov}, which 
have been observed in BEC \cite{Solitons}. 
However, bright solitons are unstable in 2D and 3D. In periodic potentials 
multidimensional discrete solitons are possible \cite{DS}, but  
they do not move in a multidimensional way, 
although the use of optical lattices has been 
proposed to move 2D and 3D discrete solitons along a 
free direction \cite{Sale}. Other interesting possibility to stabilize high dimensional solitons 
is to use Feshbach resonances to manage spatially and/or temporally the scattering length \cite{FeshbachManagment}.

Due to the DDI, a dipolar BEC is described by a NLSE with nonlocal 
cubic nonlinearity \cite{Stability,Excitations,Roton}, opening an interesting cross-disciplinary link between BEC 
and other nonlocal nonlinear media, as e.g.  
plasmas \cite{Plasma}, where the nonlocal response is induced by heating and ionization, and 
nematic liquid crystals, where it is the result of long-range 
molecular interactions \cite{Nematics}. Nonlocality plays a crucial role in the 
physics of solitons and modulation instability \cite{Bang1,Bang2,Kivshar}. 
In particular, any symmetric nonlocal nonlinear response with 
positive definite Fourier spectrum has been mathematically shown to arrest collapse in 
arbitrary dimensions \cite{Bang2}. Indeed, multidimensional solitons have been experimentally 
observed in nematic liquid crystals \cite{Nematics}. Recently we showed that under  
realistic conditions, 2D solitons may be generated in dipolar BEC \cite{Pedri05}. However, the 
anisotropic character of the DDI violates the conditions of Ref. \cite{Bang2}, and as a consequence 
a stability window occurs, rather than a stability threshold for a sufficiently large dipole strength. 
In Ref. \cite{Pedri05} we briefly studied the scattering of 2D dipolar solitons, 
which is inelastic \cite{footnote}, contrary to the 1D solitons in local NLSE, due to the lack 
of integrability \cite{Krolikowski}. However, the analysis of the inelastic scattering was largely complicated 
by the spatial overlapping of the solitons.

\begin{figure}[ht] 
\begin{center}
\includegraphics[width=5.5cm]{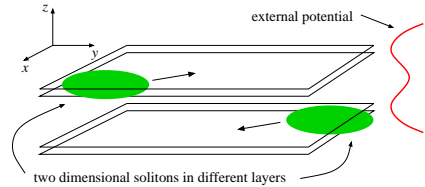}
\end{center} 
\vspace*{-0.2cm} 
\caption{Schematic representation of the system considered in this Letter.}  
\label{fi1}
\vspace*{-0.1cm}  
\end{figure}

In the following, we consider dipolar BECs placed in disconnected layers (Fig. \ref{fi1}).
For the case of short-range interactions, the two layers are independent if the 
hopping is suppressed. A remarkable consequence of the DDI is that unconnected layers
become coupled due to non-local density-density interactions, leading 
to interesting interlayer effects, as e.g. the possibility of a BEC of filaments, 
recently discussed in Ref.~\cite{Demler}. In this Letter, we analyze the rich physics 
introduced by interlayer effects in the nonlinear properties of dipolar BECs. 
In particular, we show that this interlayer interaction
leads to an effective molecular potential between fully disconnected solitons, 
allowing for a complex scattering physics between them. This novel physics includes 
inelastic fusion into excited soliton-molecules for sufficiently slow solitons, 
as well as strong symmetric and asymmetric inelastic resonances for intermediate velocities. 
In addition, we discuss a fundamentally new 2D-scattering scenario in matter-waves, 
showing that inelastic soliton spiraling similar to that 
observed in photorrefractive materials \cite{Snyder,Spiraling} is possible in dipolar BEC.
Finally we consider the scattering of 1D dipolar solitons in unconnected wires, and comment about
observability in on-going experiments.


In the following, we consider a dipolar BEC transversally 
confined in the $z$-direction by a two-well potential, with wells located at 
$z=\pm z_0$, and separated by a sufficiently large potential barrier that prevents tunneling between them.  
At each well the $z$-confinement 
is approximated by an harmonic potential of frequency $\omega_z$, whereas there is no confinement on the 
$xy$-plane. We consider in each well a BEC of $N$ particles (a more general case will be analyzed later on)
with electric dipole $d$
(the results are equally valid for magnetic dipoles) oriented in the 
$z$-direction by a sufficiently large external field, and that hence 
interact via a dipole-dipole potential:
$V_d(\vec r)=g_d (1-3\cos^2\theta)/r^3$, 
where $g_d=\alpha Nd^2/4\pi\epsilon_0$, with 
$\epsilon_0$ the vacuum permittivity, $\theta$ the angle formed by the vector 
joining the interacting particles and the dipole direction, and  
$-1/2\leq\alpha\leq 1$ a tunable parameter by means of rotating 
orienting fields \cite{Tuning}. 
At sufficiently low temperatures our system is described by  
the following two coupled NLSE with nonlocal nonlinearity \cite{Stability}:
\begin{eqnarray}
&&i\hbar\frac{\partial}{\partial t}\Psi_{j}(\vec r)=
\left [ 
-\frac{\hbar^2}{2m}\nabla^2+U_j(z)
+g|\Psi_j(\vec r)|^2 \right\delimiter 0 \nonumber \\
&&+ \left\delimiter 0  \int d\vec r' V_d(\vec r-\vec r')
(|\Psi_1(\vec r')|^2+|\Psi_{-1}(\vec r')|^2)
\right ]\Psi_j(\vec r),
\label{GPE}
\end{eqnarray}
where $j=\pm1$ is the layer-index, $\Psi_{j}$ are the wavefunctions at each well, $U_j(z)=m\omega_z^2 (z+ jz_0)^2/2$,    
$\int |\Psi_{j}(\vec r,t)|^2 d{\vec r}=1$, and 
$g=4\pi\hbar^2aN/m$ characterizes 
the contact interaction, with $a$ the $s$-wave scattering length. 
In the following we consider $a>0$, i.e. repulsive short-range interactions. 

We assume a 2D dynamics in each well. 
This approximation demands that the corresponding chemical
potential $\mu\ll\hbar\omega_z$. In that case, $\Psi_j(\vec r)\simeq \psi_j(\vec\rho)\varphi_j(z)$  
with $\varphi_j(z)$ the ground-state wave-function of the harmonic oscillator in the layer $j$.
Employing this factorization, the convolution theorem, the 
Fourier transform of the dipole-dipole potential, $\tilde V_d(k)=(4\pi/3)(3 k_z^2/k^2-1)$, 
and integrating over the $z$-direction, we arrive at a system of two coupled
2D NLSE:
\begin{eqnarray}
&&i\hbar\frac{\partial}{\partial
t}\psi_{j}=\left[-\frac{\hbar ^2}{2m}\nabla_{\rho}^2
+\frac{g|\psi_{j}|^2 }{\sqrt{2\pi}l_{z}}
+\frac{4\sqrt{\pi}g_{d}}{3\sqrt{2}l_{z}}\int
\frac{d\vec{k}_{\rho}}{(2\pi)^{2}}e^{i\vec{k}_{\rho}\vec{\rho}} \right. \nonumber \\
&&\left.  \left ( 
\tilde{n}_{j}(\vec{k}_{\rho}) F \left(k_{\rho}l_{z},0\right)
+
\tilde{n}_{-j}(\vec{k}_{\rho})
F\left(k_{\rho}l_{z},2z_0/l_{z}\right)
\right ) \right]\psi_{j}
\label{2DEQS}
\end{eqnarray}
where $l_z^2=\hbar/m\omega_z$ is the harmonic-oscillator length, $\tilde{n}_{j}$ is the Fourier transform of
$|\psi_{j}(\vec{\rho})|^{2}$
and $F(\sqrt{2}k,\sqrt{2}\lambda)=2e^{-\lambda^{2}}-
(3\sqrt{\pi}ke^{k^{2}}/2)[e^{-2k\lambda}$
erfc$(k-\lambda)+e^{2k\lambda}$erfc$(k+\lambda)]$, with erfc$(x)$ 
the complementary error function.
Using Eqs.~(\ref{2DEQS}), we study numerically the equilibrium properties and the 
dynamics of the unconnected 2D solitons.


In Ref.~\cite{Pedri05} we showed that 
if $g_d\neq 0$, a stable soliton may appear if 
$ \beta\tilde g < 3(1 +\tilde g/2\pi)/2 < -2\beta \tilde g$, where 
$\tilde g=g/\sqrt{2\pi}\hbar\omega_zl_z^3$, and $\beta=g_d/g$. 
Hence $\beta$ must be sufficiently large and negative , which 
demands tunability ($\alpha<0$). For $Na/l_z\gg 1$, stable solutions appear for 
$|\beta|>3/8\pi\simeq 0.12$. As discussed in Ref.~\cite{Pedri05}, 
the DDI may destabilize the solitons if they become 
3D. In the following, we shall restrict ourselves to the 2D regime
\cite{Pedri05}.


We introduce a variational formalism which allows us 
to study the equilibrium properties and dynamics 
of the two solitons. We consider a Gaussian Ansatz \cite{PerezGarcia}:
\begin{equation}
\psi_j(\vec \rho,t)=A\prod_{\eta=x,y}e^{-\frac{(\eta-j\eta_0)^2}{4 w_\eta^2} + i(j\eta\alpha_\eta + (\eta-j\eta_0)^2\beta_\eta)}, 
\label{Gaussian}
\end{equation}
where $A$ is the normalization factor, $\{ x_0 ,y_0\}$ is the soliton center, $w_{x,y}$ the soliton widths,
 and from the continuity equation we obtain $\alpha_\eta=m\dot\eta_0/\hbar$,
$\beta_\eta=m\dot w_\eta/2\hbar\omega_\eta$. The variables $x_0,y_0,w_x,w_y$ are time-dependent.  
The center of mass motion is an independent degree of freedom and it can be decoupled. Without loss of generality, it has not been included in the variational Ansatz. 
Introducing (\ref{Gaussian}) into the corresponding Lagrangian \cite{Pedri05}, we obtain the following set of equations of motion
\begin{equation}
m\ddot q_{i}=-\frac{\partial}{\partial q_i} U, 
\end{equation}
where $q_{\{i\}}=x_0,y_0,w_x,w_y$ are the dynamical variables, and 
\begin{equation}
U=\frac{\hbar^2}{8m} \left ( \frac{1}{w_x^2}+  \frac{1}{w_y^2} \right )
+\frac{g}{\sqrt{2\pi}8\pi w_x w_y l_z}+V, 
\label{U}
\end{equation}
is the potential energy, that includes the dipolar interaction term
\begin{eqnarray}
V&=&\frac{g_d}{12\pi^2}\int d\vec k \left(3\frac{k_z^2}{k^2} -1\right)e^{-k_z^2 l_z^2/4-k_x^2w_x^2/2-k_y^2w_y^2/2}\nonumber \\
&&(1+\cos(2k_xx_0)\cos(2k_yy_0)\cos(2k_zz_0)),
\end{eqnarray}
that couples the unconnected solitons. Hence, the problem reduces to an effective particle in the potential $U$. 
Since we have set $g_d<0$, the soliton-soliton potential is maximally 
repulsive for solitons on top of each other, becoming attractive at a given distance. 
A soliton molecule is thus possible at the minimum of $U$, which we have found 
by means of a Powell-minimization procedure, obtaining results that compare well with our 
imaginary-time simulations of Eqs.~(\ref{2DEQS}). For point-like solitons the soliton-soliton potential would be  
 $V_{point}\propto -(x_0^2-2z_0^2)/(x_0^2+z_0^2)^{5/2}$, with a minimum at $x_0=2z_0$, a value much smaller than 
that obtained from our variational or numerical calculations, showing the relevance of the 
spatial extension of the solitons. 


\begin{figure}[ht] 
\begin{center}
\includegraphics[width=5.5cm]{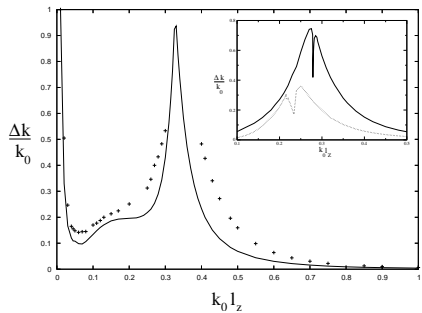}
\end{center} 
\vspace*{-0.2cm} 
\caption{Numerical (crosses) and variational (solid) results for 
$\Delta k/k_0$ ($\Delta k=k_0-k(t\rightarrow\infty)$) as a function 
of the initial momentum $k_0 l_z$, for $z_0=3l_z$, $\tilde g=200$, 
$\beta=-0.2$. Inset: Numerical results 
with $z_0=4l_z$ (solid) and $z_0=5l_z$ (dotted).}
\label{fig:1}  
\vspace*{-0.1cm}  
\end{figure}

\begin{figure}[ht] 
\begin{center}
\includegraphics[width=5.5cm]{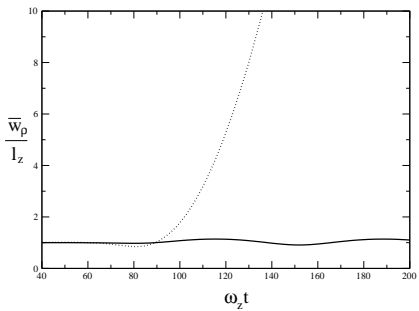}
\vspace*{-0.5cm} 
\end{center} 
\caption{Variational results for the evolution of the averaged width 
$\bar w_\rho=\sqrt{w_x^2+w_y^2}$ 
(normalized to its initial value) for soliton $1$ with $N_1=0.6N$ (dotted) 
and $2$ with $N_2=1.4N$ (solid), with $\tilde g=200$, $\beta=-0.2$, $z_0=3l_z$,  
and $k_0l_z=0.03$.}
\label{fig:difma}  
\vspace*{-0.1cm}  
\end{figure}

\begin{figure}[ht] 
\begin{center}
\includegraphics[width=5.5cm]{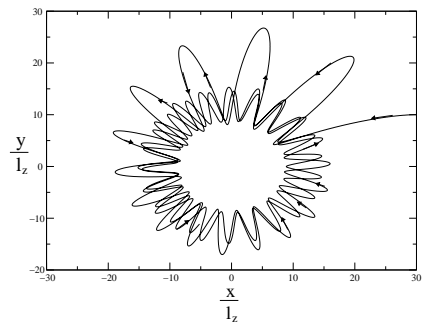}
\end{center} 
\vspace*{-0.2cm} 
\caption{Numerical result for the soliton trajectory 
during spiraling fusion in a 2D soliton scattering, for 
the case $\tilde g=200$, $\beta=-0.2$, $z_0=3l_z$, $\vec k_0 l_z=0.01 \hat x$, $x_0=30l_z$, $y_0=10l_z$.}
\label{fig:2}  
\vspace*{-0.1cm}  
\end{figure}

\begin{figure}[ht] 
\begin{center}
\includegraphics[width=5.5cm]{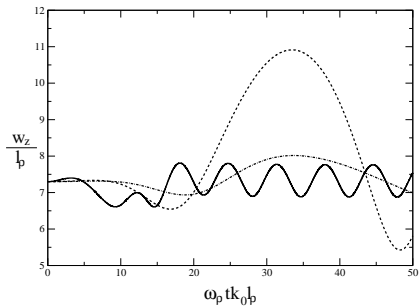}
\vspace*{-0.5cm} 
\end{center} 
\caption{Width of a 1D soliton for 
$x_0=3l_\rho$, $g/2\pi\hbar\omega_\rho l_\rho^3=25$, $\beta=0.28$, and $k_0l_\rho=0.05$ (solid), 
$0.20$ (dashed), $0.35$ (dotted-dashed). 
$\omega_\rho$ ($\l_\rho$) is the transversal oscillator frequency (length). 
The time has been re-scaled for comparison.}
\label{fig:3}  
\vspace*{-0.1cm}  
\end{figure}

In the following, we consider the scattering of solitons. We first discuss the case where the relative velocity is parallel to the vector connecting the centers of mass of the two solitons ($y_0=0$). 
We have studied the scattering for different initial
soliton velocities both by direct
numerical simulations of Eqs.~(\ref{2DEQS}) and by determining the
evolution of $\{ x_0,w_x,w_y\}$ in our variational calculation. 
Fig.~\ref{fig:1} shows the relative variation of the soliton momentum
as a function of the initial momentum. As
expected, for sufficiently large initial velocities the scattering may be
considered as elastic. For sufficiently low velocities the initial kinetic
energy of the solitons is fully transformed during the inelastic scattering 
into internal soliton energy, and the initially independent solitons 
become bounded into an excited molecular state (fusion). 

Interestingly, the inelastic losses 
do not increase monotonically for decreasing velocities, but on
the contrary show a pronounced resonant peak at intermediate velocities 
(Fig.~\ref{fig:1} and its inset). This effect is motivated by a resonant coupling to 
internal soliton modes, which leads to a dramatic enhancement of
the soliton widths after the collision, and eventually to the destruction of
the solitons. We stress that this is only possible because internal modes of the 2D soliton are 
at rather low energies, well within the inelastic regime. 
If the interlayer distance is increased, the inelastic losses are as expected reduced, but 
an even more complicated structure of resonances is then resolved (Fig.~\ref{fig:1}, inset). 

The previous analysis can be extended to asymmetric configurations, where the solitons have 
different number of particles. In this case, asymmetric inelastic scattering occurs, 
as shown in  Fig.~\ref{fig:difma}, where a resonance appears just for the solitons having the 
smaller binding energy.


Interestingly, the possibility of generating stable 2D solitons in dipolar gases allows for a
completely new scenario for soliton scattering in cold gases, namely a truly
2D scattering, in which the scattered solitons present a relative
angular momentum around the scattering center (Fig.~\ref{fig:2}). Similar scattering regimes as those discussed above
are also possible in the 2D scattering. However, the relative angular momentum during the collision 
leads, for the 2D scattering case, to a spiraling motion for sufficiently low incoming velocities, as 
a consequence of the inelastic fusion of the solitons, which stabilize in a rosetta-like
orbit around each other (Fig.~\ref{fig:2}). The 2D spiraling links the physics of dipolar BECs to that of photorrefractive 
materials, where soliton spiraling has been proposed \cite{Snyder} and experimentally observed \cite{Spiraling}.


Finally, we consider 1D BECs placed 
at neighboring 1D-sites ($x=\pm x_0$) of a 2D lattice, with the dipole oriented along the site axis. 
Following similar scaling arguments as above, 
it is possible to show that for $g>0$, a bright soliton is possible if $\beta>3/4\pi$. 
Although, of course, some 2D features are missed in 1D, it is indeed 
possible to observe inelastic processes also in 1D solitons. 
This is particularly relevant for current experiments in $^{52}$Cr, since 
no tuning is necessary for 1D solitons, easing very significantly the experimental requirements.
For $^{52}$Cr a Feshbach resonance is necessary to satisfy the previous condition, but 
Feshbach resonances are well characterized and accessible \cite{Feshbach}. In Fig.~\ref{fig:3} 
we depict an example of the dynamics of the soliton width in and out of resonance, which clearly shows  
a resonant (although non-destructive) behavior for intermediate velocities.


Summarizing, interlayer effects are a fundamentally new feature 
introduced by the DDI in dipolar gases placed in unconnected layers 
of an optical lattice. These effects may have remarkable consequences, 
as e.g. the formation of a BEC of filaments~\cite{Demler}. In this Letter, 
we discussed the rich physics introduced by interlayer effects in the nonlinear 
properties of dipolar BECs, and in particular in the 
scattering of unconnected solitons. 
The DDI induces an inelastic soliton-soliton scattering, that for low relative velocities, 
leads to the inelastic fusion into a soliton molecule. Interestingly, the inelastic 
losses do not increase monotonically for decreasing relative velocities, but on the contrary  
show strong resonances at intermediate
velocities, at which, after interacting, the soliton widths are strongly modified, 
eventually leading to soliton destruction. This effect appears, because, 
due to the relatively low excitation frequencies of the solitons, 
a resonant coupling between incoming kinetic 
energy and internal soliton modes is possible for low relative velocities
well within the inelastic regime. 
We have shown that a similar effect should be observable in 
1D geometries, where the experimental requirements may be easily fulfilled in on-going Chromium 
experiments.  Finally, we have considered the 2D scattering of dipolar solitons, a unique possibility offered
by the dipolar interactions in cold gases. We have shown that due to the combination 
of inelastic trapping and initial angular momentum a spiraling motion is possible, 
offering fascinating links to similar physics in photorrefractive materials.

\acknowledgements
Conversations with Y. Kivshar, and T. Pfau are 
acknowledged. This work was supported by the DFG (SFB-TR21, SFB407, SPP1116), 
the Minist\`ere del la Recherche (ACI Nanoscience 201), the ANR (NT05-2\_42103 and 05-Nano-008-02), and the IFRAF Institute.


\begin{thebibliography}{99}
\bibitem{footnote} Strictly speaking, we should employ the term solitary wave. For simplicity 
we use the term soliton.
\end{thebibliography}
\end{document}